\documentclass[ journal, onecolumn, draftcls]{IEEEtran} %

\usepackage{cite}      

\usepackage{graphicx}  
\usepackage{longtable}
\usepackage{psfrag}    

\usepackage{subfigure} 

\usepackage{url}       

\usepackage{stfloats}  

\usepackage{amsmath}   
\usepackage{array}
 \makeatletter
 \let\NAT@parse\undefined
 \makeatother
\usepackage{times}
\usepackage{epsfig,graphicx,color,pstricks}
\usepackage{amsmath,amssymb,amsthm,amsbsy,amsfonts,latexsym}
\usepackage{bm}
\usepackage{psfrag}
\usepackage[mathscr]{eucal}
\usepackage{cite}
\usepackage[all]{xy}
\usepackage{subfigure}
\usepackage{graphics}
\usepackage[top=0.9in, bottom=0.79in, left=0.75in, right=0.75in]{geometry}

\newtheorem{thm}{Theorem}

\newtheorem{lem}{Lemma}

\theoremstyle{remark}

\theoremstyle{definition}

\newcommand{\eu}{ {\rm e}}

\newcommand{\X}{ {\bf X}}
\newcommand{\Y}{ {\bf Y}}
\newcommand{\Z}{ {\bf Z}}
\newcommand{\V}{ {\bf V}}
\newcommand{\W}{ {\bf W}}
\newcommand{\E}{ \mathbb{E}}

\pdfoutput=1

\begin{document}
\title{Comment on the Equality Condition for the I-MMSE  Proof  of Entropy Power Inequality}
\author{
%
}

\IEEEoverridecommandlockouts                              

\author{Alex Dytso$^{*}$,  Ronit Bustin$^{**}$, H. Vincent Poor$^{*}$,  and Shlomo Shamai (Shitz)$^{**}$
\thanks{$^{*}$A. Dytso and H.V. Poor  are with  the Department of Electrical Engineering, Princeton University,  Princeton, NJ 08544, USA
(email: adytso,  poor@princeton.edu).}%
\thanks{$^{**}$R. Bustin and S. Shamai (Shitz) are with the Department of Electrical Engineering, Technion-Israel Institute of Technology, Technion City, Haifa 32000, Israel (e-mail: bustin@technion.ac.il,  sshlomo@ee.technion.ac.il).}%
  \thanks{The work of A. Dytso and H.V. Poor was supported by the National Science Foundation under Grants  CCF-1420575 and CNS-1456793. The work of S. Shamai and R.Bustin
was supported by the Unions Horizon 2020 Research and
Innovation Programme Grant 694630.
 The contents of this article are solely the responsibility of the authors and do not necessarily represent the official views of the funding agencies.}      
}

\maketitle
\begin{abstract}
The paper establishes the equality condition in the I-MMSE proof of the entropy power inequality (EPI).
This is done by establishing an exact expression for the deficit between the two sides of the EPI.
Interestingly, a necessary condition for the equality is established by making a connection to the famous Cauchy functional equation. 
\end{abstract}

The classical entropy power inequality (EPI) formulated by Shannon in  \cite{Shannon:1948} states that for two independent continuous random vectors $\V$ and $\W$ 
\begin{align}
\eu^{\frac{2}{n} h(\V+\W)} \ge  \eu^{\frac{2}{n} h(\V)}+ \eu^{\frac{2}{n} h(\W)}, \label{eq:EPI}
\end{align}
where equality in \eqref{eq:EPI} is attained if and only if  $\V$ and $\W$ are Gaussian with proportional covariances (i.e., $ {\bf K}_{W}   = c\,   {\bf K}_{V}  $ for some scalar $c>0$). 
Via the transformation
\begin{align*}
\X_1&=\frac{\V}{\sqrt{1-\alpha}} , \quad  \X_2=\frac{\W}{\sqrt{\alpha}}, \\
\alpha&= \frac{\eu^{\frac{2}{n} h(\W)}}{\eu^{\frac{2}{n} h(\V)}+ \eu^{\frac{2}{n} h(\W)}},
\end{align*}
 the EPI can be shown to be equivalent  to  Lieb's inequality \cite{lieb2002proof} 
 \begin{align}
 h \left(  \sqrt{ 1-\alpha} \X_1 +\sqrt{\alpha } \X_2 \right) \ge  \alpha h(\X_1) +(1-\alpha) h(\X_2), \, \forall  \,\alpha \in [0,1], \label{eq:Lieb}
 \end{align}
and where equality in \eqref{eq:Lieb} holds if and only if $ {\bf K}_{\X_1}   =   {\bf K}_{\X_2}$.

There are several proofs of the EPI which follow three distinct methods: using  integration
over a path of a continuous Gaussian perturbation \cite{stam1959some,blachman1965convolution,DemboCoverInfoInequalities,verdu2006simple,guo2006proof}; using the sharp version of Young's inequality  and properties of R\'enyi entropy \cite{lieb2002proof,DemboCoverInfoInequalities,blachman1965convolution}; and  using a change of variable and Kn\"{o}the's map  \cite{rioul2016yet,rioul2017optimal}.     For a comprehensive list of references and a detailed history of the EPI, the reader is referred to \cite{rioul2011information} and references therein.

As was recently pointed out in \cite{rioul2016yet} not all available proofs settle the \emph{equality} case in \eqref{eq:EPI} and \eqref{eq:Lieb}. In particular, for the class of  proofs via Gaussian perturbations, the case of equality has not yet been established in the proof  given in \cite{verdu2006simple}, which relies on the so-called I-MMSE relationship \cite{I-MMSE}.

   The goal of this paper is to close this gap by establishing the equality case in the proof of the EPI via the I-MMSE relationship. Equality is established by determining an exact expression for the deficit in \eqref{eq:Lieb} and showing that the deficit is zero if and only if $\X_1$ and $\X_2$ are Gaussian with identical covariances.

\paragraph*{Notation}
Deterministic scalar/vector quantities are denoted by lowercase normal/bold letters,
matrices 
by bold uppercase letters,
random variables 
by uppercase letters, and random vectors 
by bold uppercase letters. For a random vector $\V$ we denote the covariance matrix by $ {\bf K}_{\V}$, determinant by $|{\bf K}_{\V}|$, transpose  by  $\V^{T}$, and trace by ${\rm Tr}\{\V\}$.  
The Euclidian norm of a vector ${\bf v}$ is denoted by $\| {\bf v}\|$.  The gradient operator is denoted by $ \nabla$. The $\E[ \cdot ]$ denotes the expectation operator.
\paragraph*{Assumptions}
Throughout the paper, we assume that all random vectors treated in this work have covariance matrices with bounded entries and continuous, positive, and differentiable probability densities. Therefore, quantities such as entropies, expectations, and conditional expectations are well defined throughout the paper.   The interested reader is referred to \cite{rioul2016yet} and \cite{bobkov2015entropy} where it is shown that the set of aforementioned assumptions is sufficient to prove the EPI  in \eqref{eq:EPI}.

\section{Preliminary Results} 

In this section, we present necessary mathematical tools needed in this paper. 

%

%

The first result of this section  establishes the penalty, incurred in the minimum mean square error (MMSE), for using a sub-optimal estimator. 

\begin{lem}\label{lem:Decomp} Let  $f: \mathbb{R}^n \to \mathbb{R}^n$ be measurable and  such that $\E\left[\| f(\Y)\|^2 \right] \le \infty$. Then, 
\begin{align} 
\E \left[ \|\X-\E[\X \mid \Y] \|^2 \right]  = \E \left[ \|\X-f(\Y) \|^2 \right]  - \E \left[ \left\| f(\Y)-\E[\X \mid \Y] \right \|^2 \right]. 
\end{align}
\end{lem}
\begin{IEEEproof}
\begin{subequations}
\begin{align}
\E \left[ \left\| f(\Y)-\E\left[ \X \mid \Y\right]  \right \|^2 \right]&=\E \left[ \left\| f(\Y)-\X+\X-\E \left[\X \mid \Y \right]  \right \|^2 \right] \notag\\
&= \E \left[ \|\X-f(\Y) \|^2 \right] + \E \left[\|\X-\E\left[\X \mid \Y\right]  \|^2 \right]   + 2 \E\left[  {\rm Tr} \{(f(\Y)-\X)(\X-\E\left[ \X \mid\Y\right] )^T \} \right]  \notag  \\
&= \E \left[ \|\X-f(\Y) \|^2 \right] + \E \left[(\X-\E\left[\X \mid \Y\right] )^2 \right]   - 2\E\left[ {\rm Tr} \{\X (\X-\E\left[ \X \mid \Y\right] )^T \} \right]   \label{eq:OrthodPart1} \\
&= \E \left[\|\X-f(\Y) \|^2 \right] + \E \left[\| \X-\E\left[\X \mid \Y\right]  \|^2 \right]   - 2\E\left[ \|\X-\E\left[ \X \mid \Y\right]  \|^2 \right]  \label{eq:OrthodPart2}  \\
&= \E \left[ \|\X-f(\Y) \|^2 \right] - \E \left[ \|\X-\E\left[ \X\mid \Y\right]  \|^2 \right] ,   \notag
\end{align}
\end{subequations}
where  \eqref{eq:OrthodPart1} and \eqref{eq:OrthodPart2}  are due to the orthogonality principle. This concludes the proof. 
\end{IEEEproof}

The necessary condition for the equality in \eqref{eq:Lieb} will be shown to be a consequence of a remarkably simple, yet powerful, Cauchy functional equation. 

\begin{lem} \label{lem:CaucyFunctionalEquation}(\emph{Cauchy Functional Equation}.)  Over a space of measurable\footnote{In this paper, measurable is meant with respect to the Lebesgue measure.} functions from $\mathbb{R}^n$ to $\mathbb{R}^n$ the equation 
\begin{align}
f({\bf x}+{\bf y})=f({\bf x})+f({\bf y}),
\end{align}
 is satisfied if and only if   $f({\bf x})={\bf A} \, {\bf x}$ (i.e., is linear) for some matrix ${\bf A} \in \mathbb{R}^{n \times n}$. 
\end{lem}
\begin{IEEEproof}
See \cite[Chapter 2]{aczel1966lectures}.
\end{IEEEproof}

Cauchy functional equation has a very rich history, and the interested reader is referred to \cite{aczel1966lectures} for a comprehensive summary. 
Cauchy functional equation is used next to establish the following property of the  conditional expectation.

\begin{lem}\label{lem:condExpect} Let $\V, \W \in \mathbb{R}^n$ be independent random vectors with support and $f_1, \, f_2 :\mathbb{R}^n \to \mathbb{R}^n$ be measurable functions such that $\E\left[\| f_1(\V) \|^2 \right], \,\E \left[\| f_2(\W) \|^2 \right]<\infty$.  Then, for any $a_1,a_2 \in \mathbb{R}$
\begin{subequations}
\begin{align}
\E \left[ a_1 \,  f_1(\V)+a_2 \, f_2(\W) \mid a_1\V+a_2\W \right]=  a_1 \,  f_1(\V)+a_2 \, f_2(\W)  \text{ a.s., } \label{eq:linearityOfConditional}
\end{align}
if and only if  $f_1({\bf v})$ and $f_2({\bf w})$ are affine functions with the same slope, that is 
\begin{align}
 f_1({\bf v})= {\bf A}\, {\bf v}+{\bf b} , \  f_2({\bf w})={\bf A}\, {\bf w}+{\bf c},
\end{align}
\end{subequations}
for some  ${\bf A} \in \mathbb{R}^{n \times n}$ and ${\bf b},{\bf c} \in\mathbb{R}^n $.
\end{lem}

\begin{IEEEproof}
The proof of the sufficient condition follows trivially.  
To show the necessary condition observe that \eqref{eq:linearityOfConditional} is equivalent to identifying a set of functions  $ \{h(\cdot ) \}$  for which 
\begin{align}
h( a_1 \W+ a_2 \V)=a_1 \,  f_1(\V)+a_2 \, f_2(\W).
\end{align}
Since $\V$ and $\W$ are fully supported,
we have that
\begin{align}
h( a_1 {\bf v}+a_2{\bf w})=a_1 \, f_1({\bf v})+a_2 \, f_2({\bf w}), \label{eq:COnditionBBLL}
\end{align}
 for all $({\bf v},{\bf w})$. In particular,
\begin{subequations}
\begin{align}
h( a_1{\bf v})=a_1 \, f_1({\bf v})+a_2 \, f_2(0),\\
h( a_2{\bf w})=a_1 \, f_1(0)+a_2 \, f_2({\bf w}).
\end{align}
\label{eq:linearSystem}
\end{subequations} 
Therefore, by adding the two equations in \eqref{eq:linearSystem} and using \eqref{eq:COnditionBBLL}, we arrive at
\begin{align}
h(a_1{\bf v})+h(a_2{\bf w})&=a_1 \, f_1({\bf v})+ a_2 \, f_2(0)+a_1 \, f_1(0)+a_2 \, f_2({\bf w})= h( a_1{\bf v}+a_2{\bf w})+h(0). \label{eq:CauchyEquivalent}
\end{align}
Next, by letting   $f({\bf x})=h({\bf x})-h(0)$,  it is not difficult to see that \eqref{eq:CauchyEquivalent}  corresponds to Cauchy functional equation in Lemma~\ref{lem:CaucyFunctionalEquation}.
As a result, we concluded that $h(\cdot)$ is an affine function 
\begin{align}
h({\bf x} )= {\bf A} {\bf x} +{\bf a}, \label{eq:hISaffine}
\end{align}
for some ${\bf A} \in \mathbb{R}^{n \times n}$ and ${\bf a} \in \mathbb{R}^n$. Finally, \eqref{eq:hISaffine} and  \eqref{eq:linearSystem} imply that  functions $f_1(\cdot)$ and $f_2(\cdot)$ are also affine with the same slope.  This concludes the proof. 
\end{IEEEproof}

The following well-known property of the conditional expectation will be useful in manipulating some of our expressions.

\begin{lem}\label{lem:ToweringProp} (\emph{Smoothing or Towering Property of the Conditional Expectation.}) Let   sigma algebras $\mathcal{G}_1,\mathcal{G}_2$ be such that $\mathcal{G}_1 \subset \mathcal{G}_2$. Then,
\begin{subequations}
\begin{align}
\E \left[  \E \left[ \X  \mid \mathcal{G}_2 \right]  \mid  \mathcal{G}_1 \right]=\E \left[ \X  \mid \mathcal{G}_1 \right] \text{ a.s.}
\end{align}
In particular,  for $(\X,\W,\V)$
\begin{align}
\E \left[  \E \left[ \X  \mid \W,\V  \right]  \mid  \W+\V \right]=\E \left[ \X  \mid \W+\V \right]  \text{ a.s.}
\end{align}
\end{subequations}
\end{lem}
\begin{IEEEproof}
See \cite[Chapter 10]{resnick2013probability}.
\end{IEEEproof}

The key step of the proof would be to establish that the equality holds if and only if a certain conditional expectation is a linear or an affine function. The following result shows that  the conditional expectation is an affine function if and only if the input random variable is Gaussian.

\begin{lem} \label{lem:GaussianANDLinearity}  Let  $\Y= \, \X+\Z$ where $\X$ and $\Z \sim \mathcal{N}(0,{\bf I})$ are independent. Then,
\begin{subequations}
\begin{align}
\E[\X \mid \Y]=  {\bf A}  \Y+ {\bf v}  \text{ a.s.},
\end{align}
 if and only if  $\X ~\sim \mathcal{N}(\mu_{\X}, {\bf K}_{\X})$  such that
 \begin{align}
 {\bf A} = {\bf K}_{\X \Y} {\bf K}_{\Y}^{-1}= {\bf K}_{\X}  ( {\bf I} + {\bf K}_{\X})^{-1},  \, {\bf v}=({\bf I}- {\bf A} )   \mu_\X . 
 \end{align}
 \end{subequations}
\end{lem} 
\begin{IEEEproof}
Lemma~\ref{lem:GaussianANDLinearity}  is a well known result from estimation theory and the details of the proof can be found in \cite{poor2013introduction}. Here, we only a give sketch of the proof of the necessary condition.   To show the necessary condition, one must show that, in the MMSE sense, linear estimators are only optimal for  Gaussian random vectors.  For simplicity, we only look at the zero mean (i.e., $ \mu_X=\E[X]=0$ and $v=0$) and the scalar case. 
Let  $f(Y)=\eu^{-itY}$, and let $aY$ be an estimator that we claim to be optimal with $a=\frac{E[X Y]}{ E[Y^2]}$.   Then, by the orthogonality principle we have that
\begin{subequations}
\begin{align}
0&=\E \left[ (X -a Y) \, \eu^{-itY} \right] \notag\\
&=\E \left[  \left( (1-a ) X-Z  \right)\eu^{-itY} \right] \notag\\
&= (1-a) \E \left[  X \eu^{-itY} \right]-\E \left[  Z\eu^{-itY} \right] \notag\\
&= (1-a) \E \left[  X \eu^{-it  X} \right] \, \E \left[  \eu^{-itZ} \right]-\E \left[  Z\eu^{-itZ} \right] \, \E \left[  \eu^{-itX} \right] \label{eq:GaunssLinearity:Independnce}\\
&= (1-a) i  \phi_X^{'}(  \,t)  \phi_Z(t)-i \phi_Z^{'}(t)  \phi_X( \,t) \label{eq:GaunssLinearity:Char} \\
&= (1-a) \,  i \,\phi_X^{'}(  t) \, \eu^{-\frac{t^2}{2}}-i \, t  \, \eu^{-\frac{t^2}{2}}  \phi_X(t), \label{eq:GaunssLinearity:Final}
\end{align}
\end{subequations}
where \eqref{eq:GaunssLinearity:Independnce} follows by the independence of $X$ and $Z$,  and \eqref{eq:GaunssLinearity:Char} follows by the derivative expression $\phi_X^{'}(t)= - i \,\E \left[    X \eu^{-itX}\right]$ (the derivative expression holds since by assumption $\E[X^2]<\infty)$. 

Therefore, from \eqref{eq:GaunssLinearity:Final} we have a differential equation of the form
\begin{align}
(1-a) \,\phi_X^{'}(  t) = t \, \phi_X(  t).  \label{eq:diffEquation}
\end{align}
The only nontrivial solution to the  differential equation in \eqref{eq:diffEquation} is given by  the Gaussian distribution with the characteristic function given by $\phi_X(t)=\eu^{- (a-1) \frac{ t^2}{2} }$. This concludes the proof. 
\end{IEEEproof}

We define the \emph{score function} and the \emph{Fisher information} of a continuous random vector $\X$ with the probability density function $f_{\X}({\bf x})$ as 
\begin{subequations}
\begin{align}
\rho_{\X}({\bf x} )&= \nabla_{\bf x} \log \left(f_{\X}({\bf x}) \right),\\
J(\X)&=  \E \left[ \rho_{\X}(\X )^T \rho_{\X}(\X ) \right].
\end{align}
\label{eq:FisherInformation}
\end{subequations}

For the Gaussian noise channel, the  score function of the output can be related  to the conditional expectation.

\begin{lem}\label{lem:scoreFunction} Let $\Y=\sqrt{\gamma} \, \X+\Z$  where $\X$ and $\Z \sim \mathcal{N}( 0,{\bf I})$ are independent. Then,
\begin{align}
\rho_{\Y}({\bf \Y} )= \sqrt{\gamma} \, \E[\X \mid \Y] -\Y \text{ a.s.}
\end{align}
\end{lem}
\begin{IEEEproof}
See \cite[Eq.(56)]{I-MMSE}.
\end{IEEEproof}

We conclude this section by giving an expression for the differential entropy in terms of an integral of the MMSE which is a consequence of the  I-MMSE relationship in \cite{I-MMSE}.

\begin{lem}\label{lem:I-MMSEforEntropy} For every continuous random vector $\X \in \mathbb{R}^n$,
\begin{subequations}
\begin{align}
h(\X) =\frac{1}{2} \int_0^\infty  \E \left[ \| \X-\E[\X \mid \Y_\gamma] \|^2 \right] -\frac{n}{2 \pi \eu +\gamma} d \gamma,
\end{align}
as long as 
\begin{align}
\lim_{t \to 0}h(\X+t \Z)=h(\X),
\end{align}
\end{subequations}
where $ \Y_\gamma=\sqrt{\gamma} \, \X+\Z$ and $\X$ is independent of  $\Z\sim \mathcal{N}(0, {\bf I})$.
\end{lem}

\section{Main Results}

The first main result of this section, which is a refinement of the bound in \cite{verdu2006simple}, establishes an exact expression for the deficit in \eqref{eq:Lieb}.

\begin{thm} For any independent continuous random vectors $\X_1, \X_2 \in \mathbb{R}^n$ and any $\alpha \in [0,1]$
\begin{subequations}
\begin{align}
h (\sqrt{1-\alpha} \X_1 +\sqrt{\alpha} \X_2) =  (1-\alpha) h(\X_1) + \alpha h(\X_2)   + \Delta(\X_1 \| \X_2), 
\end{align}
where 
\begin{align}
\Delta(\X_1 \| \X_2)&= \frac{1}{2} \int_0^\infty \E \left[ \left \|  \E[ \X \mid \sqrt{1-\alpha} \Y_{1,\gamma}+\sqrt{\alpha} \Y_{2,\gamma}] -  \E[ \X \mid \Y_{1,\gamma}, \Y_{2,\gamma}] \right \|^2 \right] d \gamma, \label{eq:DefinitionOfDistance}\\
\X&=\sqrt{1-\alpha} \X_1 +\sqrt{\alpha} \X_2, \label{eq:DefintionOfCombX1X2}\\
\Y_{1,\gamma}&=  \sqrt{\gamma} \X_1 +\Z_1,\\
\Y_{2,\gamma}&=  \sqrt{\gamma} \X_1 +\Z_2,
\end{align}
where $\Z_1\sim \mathcal{N}(0, {\bf I}), \, \Z_2\sim \mathcal{N}(0, {\bf I})$ and where $(\X_1,\X_2,\Z_1,\Z_2)$ are mutually independent. 
\end{subequations} 

\end{thm}
\begin{IEEEproof}
According to Lemma~\ref{lem:I-MMSEforEntropy}, the entropy of a random vector $\X \in \mathbb{R}^n$, defined in \eqref{eq:DefintionOfCombX1X2}, is given by
\begin{align}
h(\X)&= \frac{1}{2} \int_0^\infty  \E[ \| \X-\E[\X \mid \Y_\gamma] \|^2] -\frac{n}{2 \pi \eu +\gamma} d \gamma \notag\\
& = \frac{1}{2} \int_0^\infty  \E \left[ \| \X-\E \left[\X \mid \Y_{1,\gamma},\Y_{2,\gamma} \right] \|^2  \right] + \E \left[  \| \E[\X \mid \Y_{\gamma}]-\E[ \X \mid \Y_{1,\gamma},\Y_{2,\gamma} ] \|^2 \right]  -\frac{n}{2 \pi \eu +\gamma} d \gamma, \label{eq:entropyX}
\end{align}
where the last step follows   by taking
\begin{align}
f( \Y_{1,\gamma},\Y_{2,\gamma} )&= \E \left[\X \mid \sqrt{1-\alpha} \Y_{1,\gamma}+\sqrt{\alpha} \Y_{2,\gamma} \right],
\end{align}
in Lemma~\ref{lem:Decomp}.
 
Next,  by the  mutual independence of $(\X_1,\X_2,\Z_1,\Z_2)$, the first expectation in \eqref{eq:entropyX} reduces to
\begin{align}
 \E \left[ \| \X-\E[\X\mid \Y_{1,\gamma},\Y_{2,\gamma} ] \|^2  \right]= (1-\alpha) \, \E \left[ \| \X_1-\E[\X_1 \mid \Y_{1,\gamma}]\|^2 \right] +\alpha \,  \E \left[\|\X_2-\E[\X_2 \mid \Y_{2,\gamma}]\|^2\right]. \label{eq:DecompositionOfMMSE} 
\end{align} 
Finally, by combining \eqref{eq:entropyX} and \eqref{eq:DecompositionOfMMSE} we arrive at 
\begin{align*}
h(\X)&=  \frac{1}{2} \int_0^\infty   (1-\alpha)\, \E \left[ \| \X_1- \E[\X_1 \mid \Y_{1,\gamma}] \|^2  \right] +  \alpha \, \E \left[ \| \X_2- \E[\X_2|\Y_{2,\gamma}] \|^2  \right]   \notag
\\&+ \E \left[ \|\E[\X \mid \Y_{\gamma}]-\E[\X \mid \Y_{1,\gamma},\Y_{2,\gamma} ] \|^2 \right]  -\frac{n}{2 \pi \eu +\gamma} d \gamma\\
&= (1- \alpha)\, h(\X_1) + \alpha \,h(\X_2) +\Delta(\X_1 \| \X_2),
\end{align*} 
where $\Delta(\X_1 \| \X_2)$ is defined in \eqref{eq:DefinitionOfDistance}. 
This concludes the proof. 
\end{IEEEproof}

Clearly,  $\Delta(\X_1 \| \X_2)$  in \eqref{eq:DefinitionOfDistance} is a non-negative quantity which leads to Lieb's inequality in \eqref{eq:Lieb}.

\subsection{On the Equality Condition} 
The following result establishes necessary and sufficient conditions for the equality in  \eqref{eq:Lieb} and gives several equivalent statemens for the equality.  

\begin{thm}\label{thm:SuffAndNess} The following statements are equivalent:
\begin{subequations}
    \begin{align}
\Delta(\X_1 \| \X_2) &= 0, \label{eq:ConditionForEqualityOriginal}\\
\E[ \X \mid \sqrt{1-\alpha} \Y_{1,\gamma}+\sqrt{\alpha} \Y_{2,\gamma}] &= \E[ \X \mid  \Y_{1,\gamma}, \Y_{2,\gamma}]   \text{ a.s.},   \label{eq:ConditionForEqualityConditionalExpect}\\
 \E \left[  \sqrt{1-\alpha} \, \E[\X_1 \mid \Y_{1,\gamma}] +\sqrt{\alpha} \,  \E[\X_2 \mid \Y_{2,\gamma}] \mid \sqrt{1-\alpha} \Y_{1,\gamma}+\sqrt{\alpha} \Y_{2,\gamma} \right] &= \sqrt{1-\alpha} \, \E[\X_1 \mid \Y_{1,\gamma}] +\sqrt{\alpha} \,  \E[\X_2 \mid \Y_{2,\gamma}]  \text{ a.s.},   \label{eq:ConditionForEqualityConditionalExpectTowering}\\
  \E [  \sqrt{1-\alpha} \rho_{\Y_{1,\gamma}} (\Y_{1,\gamma} )+ \sqrt{\alpha} \rho_{\Y_{2,\gamma}} (\Y_{2,\gamma} )   \mid \sqrt{1-\alpha} \Y_{1,\gamma}+\sqrt{\alpha} \Y_{2,\gamma} ] &=   \sqrt{1-\alpha} \rho_{\Y_{1,\gamma}} (\Y_{1,\gamma} )+ \sqrt{\alpha} \rho_{\Y_{2,\gamma}} (\Y_{2,\gamma} ) \, \text{ a.s.}, \label{eq:ConditionScoreFunction}\\
  \rho_{\sqrt{1-\alpha} \Y_{1,\gamma} + \sqrt{\alpha} \Y_{2,\gamma} }(\sqrt{1-\alpha} \Y_{1,\gamma} + \sqrt{\alpha} \Y_{2,\gamma} )&=\sqrt{1-\alpha} \rho_{\Y_{1,\gamma}} (\Y_{1,\gamma} )+ \sqrt{\alpha} \rho_{\Y_{2,\gamma}} (\Y_{2,\gamma} ) \, \text{ a.s.}, \label{eq:ConditionScoreFunction2}\\
  (1-\alpha) \, J \left(\Y_{1,\gamma})+\alpha \, J(\Y_{2,\gamma} \right)&=  J \left(\sqrt{1-\alpha} \Y_{1,\gamma} + \sqrt{\alpha} \Y_{2,\gamma} \right ). \label{eq:FisherInformationEquality}
\end{align}
\label{eq:ConditionsForEquality}
\end{subequations}
Moreover, equality in \eqref{eq:ConditionsForEquality} holds if and only if  $\X_1$ and $\X_2$ are  Gaussian with identical covariances.  
\end{thm} 

\begin{IEEEproof}

From \eqref{eq:DefinitionOfDistance} it is immediate that  $\Delta(\X_1 \| \X_2) = 0$  if and only if
\begin{align}
\E[ \X \mid \sqrt{1-\alpha} \Y_{1,\gamma}+\sqrt{\alpha} \Y_{2,\gamma}] = \E[ \X \mid  \Y_{1,\gamma}, \Y_{2,\gamma}]  \text{ a.s.} 
\end{align} 
This shows equivalence between \eqref{eq:ConditionForEqualityOriginal}  and \eqref{eq:ConditionForEqualityConditionalExpect}. 

The  equivalence between  \eqref{eq:ConditionForEqualityConditionalExpect} and \eqref{eq:ConditionForEqualityConditionalExpectTowering}  follows from the towering property in Lemma~\ref{lem:ToweringProp}
\begin{align}
\E \left[  \sqrt{1-\alpha} \, \E[\X_1 \mid \Y_{1,\gamma}] +\sqrt{\alpha} \,  \E[\X_2 \mid \Y_{2,\gamma}] \mid \sqrt{1-\alpha} \Y_{1,\gamma}+\sqrt{\alpha} \Y_{2,\gamma} \right] &= \E \left[  \E[ \X \mid  \Y_{1,\gamma}, \Y_{2,\gamma}] \mid \sqrt{1-\alpha} \Y_{1,\gamma}+\sqrt{\alpha} \Y_{2,\gamma} \right] \notag\\
&= \E \left[  \X  \mid \sqrt{1-\alpha} \Y_{1,\gamma}+\sqrt{\alpha} \Y_{2,\gamma} \right].
\end{align}

Showing  equivalence between \eqref{eq:ConditionScoreFunction}, \eqref{eq:ConditionScoreFunction2} and \eqref{eq:FisherInformationEquality}  is deferred to Appendix~\ref{sec:EquaivalenceContinued}.

Next, we show that \eqref{eq:ConditionsForEquality} is satisfied if and only if $\X_1$ and $\X_2$ are Gaussian random vectors with identical covariances.   The sufficient condition follows by noting that  if $\X_1\sim \mathcal{N}(0, {\bf K}_{\X_1})$ and $\X_2\sim \mathcal{N}(0,{\bf K}_{\X_2})$, then the estimators are linear and  are given by 
\begin{align*}
\E[ \X \mid \sqrt{1-\alpha} \Y_{1,\gamma}+\sqrt{\alpha} \Y_{2,\gamma}]&=     {\bf K}_{\X \, \left( \sqrt{1-\alpha} \Y_{1,\gamma}+\sqrt{\alpha} \Y_{2,\gamma} \right)}  \,{\bf K}_{\sqrt{1-\alpha} \Y_{1,\gamma}+\sqrt{\alpha} \Y_{2,\gamma}}^{-1} \,  \left( \sqrt{1-\alpha} \Y_{1,\gamma}+\sqrt{\alpha} \Y_{2,\gamma} \right),\\
\E[ \X_1 \mid \Y_{1,\gamma}] &=   {\bf K}_{\X_1 \, \Y_{1,\gamma} } \, {\bf K}_{\Y_{1,\gamma}}^{-1} \,   \Y_{1,\gamma},\\
\E[ \X_2 \mid \Y_{2,\gamma}] &=   {\bf K}_{\X_2 \, \Y_{2,\gamma} } \, {\bf K}_{\Y_{2,\gamma}}^{-1} \,   \Y_{2,\gamma}.
\end{align*}
Therefore,  the equality condition  in \eqref{eq:ConditionForEqualityConditionalExpect} holds only if 
\begin{subequations}
\begin{align}
 {\bf K}_{\X \, \left( \sqrt{1-\alpha} \Y_{1,\gamma}+\sqrt{\alpha} \Y_{2,\gamma} \right)}  \,{\bf K}_{\sqrt{1-\alpha} \Y_{1,\gamma}+\sqrt{\alpha} \Y_{2,\gamma}}^{-1}=  {\bf K}_{\X_1 \, \Y_{1,\gamma} } \, {\bf K}_{\Y_{1,\gamma}}^{-1} \, , \\
  {\bf K}_{\X \, \left( \sqrt{1-\alpha} \Y_{1,\gamma}+\sqrt{\alpha} \Y_{2,\gamma} \right)}  \,{\bf K}_{\sqrt{1-\alpha} \Y_{1,\gamma}+\sqrt{\alpha} \Y_{2,\gamma}}^{-1}=  {\bf K}_{\X_2 \, \Y_{2,\gamma} } \, {\bf K}_{\Y_{2,\gamma}}^{-1}.
\end{align} 
\label{eq:CovarienceCondtion}
\end{subequations}
With a small amount of algebra it is not difficult to show that the equality in \eqref{eq:CovarienceCondtion} holds only if $ {\bf K}_{\X_1 }={\bf K}_{\X_2 }$. 

The necessary condition follows by  letting  $f_1(\Y_{1,\gamma})=  \, \E[\X_1 \mid \Y_{1,\gamma}]$ and $f_2(\Y_{2,\gamma})=  \E[\X_2 \mid \Y_{2,\gamma}]$, in which case the condition in \eqref{eq:ConditionForEqualityConditionalExpectTowering} reduces to 
\begin{align}
\E[ \sqrt{1-\alpha} \,f_1(\Y_{1,\gamma})+ \sqrt{\alpha} \,  f_2(\Y_{2,\gamma}) \mid  \sqrt{1-\alpha}\Y_{1,\gamma}+ \sqrt{\alpha}\Y_{2,\gamma}]=   \sqrt{1-\alpha} \,f_1(\Y_{1,\gamma})+ \sqrt{\alpha} \,  f_2(\Y_{2,\gamma}). \label{eq:labelCondtiondafda}
\end{align}
According to Lemma~\ref{lem:condExpect}  equality in   \eqref{eq:labelCondtiondafda} implies that  $ f_1(\Y_{1,\gamma})$ and $f_2(\Y_{2,\gamma})$ (or  $ \E[\X_1 \mid \Y_{1,\gamma}]$ and $\E[\X_2 \mid \Y_{2,\gamma}]$) are affine functions with the same slope. In other words, the conditional expectations are given by 
\begin{align*}
\E[\X_1 \mid \Y_{1,\gamma}]= {\bf A}  \Y_{1,\gamma} +{\bf b},\\
\E[\X_2 \mid \Y_{2,\gamma}]= {\bf A}  \Y_{2,\gamma} +{\bf c}.
\end{align*} 
Moreover, by Lemma~\ref{lem:GaussianANDLinearity} the  linearity of conditional expectations implies that $\X_1$ and $\X_2$ are Gaussian random vectors such that
\begin{align}
{\bf A} =  \sqrt{\gamma} \,  {\bf K}_{\X_1} ( {\bf I} + \gamma \, {\bf K}_{\X_1}  )^{-1}= \sqrt{\gamma} \,  {\bf K}_{\X_2} ( {\bf I} + \gamma \, {\bf K}_{\X_2}  )^{-1}. \label{eq:CoveriencesInquality} 
\end{align}
From \eqref{eq:CoveriencesInquality}, it is evident that  $\X_1$ and $\X_2$ have identical covariances.  This concludes the proof. 
\end{IEEEproof} 

\section{Concluding Remark}
In this work, we have established the equality condition for the I-MMSE proof of the EPI. Theorem~\ref{thm:SuffAndNess} also establishes an equality condition for  the following  Fisher information inequality
\begin{align}
(1-\alpha) J(\X_1) + \alpha J(\X_2) \ge J (\sqrt{1-\alpha} \X_1 +\sqrt{\alpha} \X_2).\label{eq:FisherInequality}
\end{align}
This should come as no surprise since the inequality in \eqref{eq:FisherInequality} is a key to establishing the proof of the EPI via DeBruijn's identity \cite{stam1959some,blachman1965convolution,DemboCoverInfoInequalities}.
The equality condition in \eqref{eq:FisherInequality} was previously established in \cite{carlen1991entropy}  by showing that a certain differential equation is satisfied only by Gaussian densities, and in \cite{madiman2007generalized}  by checking the equality case of the \emph{Variance Drop} inequality. In contrast, our proof relies on Cauchy functional equation and towering property of the conditional expectation.

It is also interesting to observe that  the expression for the deficit 
\begin{align}
2 \, \Delta(\X_1 \| \X_2)&= \int_0^\infty \E \left[ \left \|  \E[ \X \mid \sqrt{1-\alpha} \Y_{1,\gamma}+\sqrt{\alpha} \Y_{2,\gamma}] -  \E[ \X \mid \Y_{1,\gamma}, \Y_{2,\gamma}] \right \|^2 \right] d \gamma\\
&=  \int_0^\infty  \frac{1}{\gamma} \E \left[ \left \|  \rho_{\sqrt{1-\alpha} \Y_{1,\gamma}+\sqrt{\alpha} \Y_{2,\gamma}}(\sqrt{1-\alpha} \Y_{1,\gamma}+\sqrt{\alpha} \Y_{2,\gamma}) - \sqrt{1-\alpha}\rho_{ \Y_{1,\gamma}}(\Y_{1,\gamma})-\sqrt{\alpha} \rho_{\Y_{2,\gamma}}(\Y_{2,\gamma})   \right \|^2 \right] d \gamma,
\end{align}
is closely related to the mismatched representation of the relative entropy \cite{VerduMismatchedMSE} 
\begin{align}
2 \,D(P\| Q) &=  \int_0^\infty  \E_{P} \left[ \left \| \E_{P}[ \X_1 \mid \Y_{\gamma}] -  \E_{Q}[ \X_2 \mid \Y_{\gamma}] \right \|^2 \right]  d \gamma \\
&= \int_0^\infty   \frac{1}{\gamma } I( P_{\Y_{\gamma}} \| Q_{\Y_\gamma}) d\gamma,
\end{align}
where $\X_1 \sim P, \, \X_2 \sim Q$, and $\sqrt{\gamma} \,\X_1+\Z \sim P_{\Y_{\gamma}},\,  \sqrt{\gamma} \, \X_2+\Z \sim  Q_{\Y_\gamma}$, and where 
\begin{align}
 I( P_{\Y_\gamma} \| Q_{\Y_\gamma}) =  \E_{P} \left[ \left \| \rho_{P_{\Y_\gamma}} ( \Y_{\gamma}) -  \rho_{Q_{\Y_\gamma} }( \Y_{\gamma}) \right \|^2 \right], 
\end{align} 
is the \emph{relative Fisher information distance}.  

Moreover, in view of the exact characterization of $\Delta(\X_1 \| \X_2)$  in \eqref{eq:DefinitionOfDistance}, it would be interesting to explore connections to the work in \cite{courtade2016wasserstein}.
The authors of   \cite{courtade2016wasserstein}  provided lower bounds on $\Delta(\X_1 \| \X_2)$,    for log-concave densities in terms of Wasserstein distance, by  using the lower bound on $\Delta(\X_1 \| \X_2)$ from \cite{rioul2016yet}.

\begin{appendices}
\section{Proof of equivalence between  \eqref{eq:ConditionScoreFunction}, \eqref{eq:ConditionScoreFunction2}, and \eqref{eq:FisherInformationEquality}} 
\label{sec:EquaivalenceContinued}
The equivalence between \eqref{eq:ConditionForEqualityConditionalExpectTowering} and \eqref{eq:ConditionScoreFunction} follows from Lemma~\ref{lem:scoreFunction}.

Next we show the equivalence between \eqref{eq:ConditionScoreFunction} and \eqref{eq:ConditionScoreFunction2}. Using Lemma~\ref{lem:scoreFunction} the score function can be written as
\begin{align}
\rho_{\sqrt{1-\alpha} \Y_{1,\gamma} + \sqrt{\alpha} \Y_{2,\gamma} }(\sqrt{1-\alpha} \Y_{1,\gamma} + \sqrt{\alpha} \Y_{2,\gamma} )= \sqrt{\gamma}  \, \E \left[ \X \mid \sqrt{1-\alpha} \Y_{1,\gamma} + \sqrt{\alpha} \Y_{2,\gamma} \right] -\left( \sqrt{1-\alpha} \Y_{1,\gamma} + \sqrt{\alpha} \Y_{2,\gamma}\right). \label{eq:DecompostionScoreFunction}
\end{align}
Next, by the towering property of the conditional expectation in Lemma~\ref{lem:ToweringProp}
\begin{align}
&\sqrt{\gamma} \,\E \left[ \X \mid \sqrt{1-\alpha} \Y_{1,\gamma} + \sqrt{\alpha} \Y_{2,\gamma} \right] \notag\\
 &=  \sqrt{\gamma} \, \E \left[ \E\left[ \X \mid  \Y_{1,\gamma} ,\Y_{2,\gamma} \right]  \mid \sqrt{1-\alpha} \Y_{1,\gamma} + \sqrt{\alpha} \Y_{2,\gamma} \right] \notag \\
&= \sqrt{\gamma} \, \E \left[ \sqrt{1-\alpha} \E\left[ \X_1 \mid  \Y_{1,\gamma}  \right]+\sqrt{\alpha}  \E\left[ \X_2 \mid  \Y_{2,\gamma}  \right]  \mid \sqrt{1-\alpha} \Y_{1,\gamma} + \sqrt{\alpha} \Y_{2,\gamma} \right] \notag\\
&=   \E \left[ \sqrt{1-\alpha} \left( \rho_{\Y_{1,\gamma}}(\Y_{1,\gamma})+\Y_{1,\gamma} \right)+\sqrt{\alpha} \left( \rho_{\Y_{2,\gamma}}(\Y_{2,\gamma})+\Y_{2,\gamma} \right)  \mid \sqrt{1-\alpha} \Y_{1,\gamma} + \sqrt{\alpha} \Y_{2,\gamma} \right] \label{eq:ScoreFunctionIdentity}
\end{align}
where the last step follows Lemma~\ref{lem:scoreFunction}. 
Putting equation \eqref{eq:DecompostionScoreFunction} and \eqref{eq:ScoreFunctionIdentity} together we arrive at
\begin{align}
\rho_{\sqrt{1-\alpha} \Y_{1,\gamma} + \sqrt{\alpha} \Y_{2,\gamma} }(\sqrt{1-\alpha} \Y_{1,\gamma} + \sqrt{\alpha} \Y_{2,\gamma} )=   \E \left[  \sqrt{1-\alpha} \rho_{\Y_{1,\gamma}}(\Y_{1,\gamma})+ \sqrt{\alpha}\rho_{\Y_{2,\gamma}}(\Y_{2,\gamma}) \mid \sqrt{1-\alpha} \Y_{1,\gamma} + \sqrt{\alpha} \Y_{2,\gamma} \right], \label{eq:scoreFunctionConvolutionIdentity}
\end{align}
which establishes equivalence between \eqref{eq:ConditionScoreFunction} and \eqref{eq:ConditionScoreFunction2}. The expression in \eqref{eq:scoreFunctionConvolutionIdentity} is sometimes called  a convolution identity of the score function \cite{madiman2007generalized}.

The equivalence between \eqref{eq:ConditionScoreFunction}  and \eqref{eq:FisherInformationEquality}  follows from \eqref{eq:scoreFunctionConvolutionIdentity} and the definition of Fisher's information in \eqref{eq:FisherInformation}.
This concludes the proof.

\end{appendices}

\bibliography{refs}

\begin{thebibliography}{10}
\providecommand{\url}[1]{#1}
\csname url@samestyle\endcsname
\providecommand{\newblock}{\relax}
\providecommand{\bibinfo}[2]{#2}
\providecommand{\BIBentrySTDinterwordspacing}{\spaceskip=0pt\relax}
\providecommand{\BIBentryALTinterwordstretchfactor}{4}
\providecommand{\BIBentryALTinterwordspacing}{\spaceskip=\fontdimen2\font plus
\BIBentryALTinterwordstretchfactor\fontdimen3\font minus
  \fontdimen4\font\relax}
\providecommand{\BIBforeignlanguage}[2]{{%
\expandafter\ifx\csname l@#1\endcsname\relax
\typeout{** WARNING: IEEEtran.bst: No hyphenation pattern has been}%
\typeout{** loaded for the language `#1'. Using the pattern for}%
\typeout{** the default language instead.}%
\else
\language=\csname l@#1\endcsname
\fi
#2}}
\providecommand{\BIBdecl}{\relax}
\BIBdecl

\bibitem{Shannon:1948}
C.~Shannon, ``A mathematical theory of communication,'' \emph{Bell Syst. Tech.
  J.}, vol.~27, no. 379-423, 623-656, Jul., Oct. 1948.

\bibitem{lieb2002proof}
E.~H. Lieb, ``Proof of an entropy conjecture of {W}ehrl,'' in
  \emph{Inequalities}.\hskip 1em plus 0.5em minus 0.4em\relax Springer, 2002,
  pp. 359--365.

\bibitem{stam1959some}
A.~Stam, ``Some inequalities satisfied by the quantities of information of
  {F}isher and {S}hannon,'' \emph{Information and Control}, vol.~2, no.~2, pp.
  101--112, 1959.

\bibitem{blachman1965convolution}
N.~Blachman, ``The convolution inequality for entropy powers,'' \emph{IEEE
  Trans. Inf. Theory}, vol.~11, no.~2, pp. 267--271, 1965.

\bibitem{DemboCoverInfoInequalities}
A.~Dembo, T.~Cover, and J.~Thomas, ``Information theoretic inequalities,''
  \emph{IEEE Trans. Inf. Theory}, vol.~37, no.~6, pp. 1501--1518, Nov 1991.

\bibitem{verdu2006simple}
S.~Verd{\'u} and D.~Guo, ``A simple proof of the entropy-power inequality,''
  \emph{IEEE Trans. Inf. Theory}, vol.~52, no.~5, pp. 2165--2166, 2006.

\bibitem{guo2006proof}
D.~Guo, S.~Shamai, and S.~Verd{\'u}, ``Proof of entropy power inequalities via
  {MMSE},'' in \emph{Proc. IEEE Int. Symp. Inf. Theory}.\hskip 1em plus 0.5em
  minus 0.4em\relax IEEE, 2006, pp. 1011--1015.

\bibitem{rioul2016yet}
O.~Rioul, ``Yet another proof of the entropy power inequality,'' \emph{arXiv
  preprint arXiv:1606.05969}, 2016.

\bibitem{rioul2017optimal}
------, ``Optimal transport to the entropy-power inequality and a reverse
  inequality,'' \emph{arXiv preprint arXiv:1701.08534}, 2017.

\bibitem{rioul2011information}
------, ``Information theoretic proofs of entropy power inequalities,''
  \emph{IEEE Trans. Inf. Theory}, vol.~57, no.~1, pp. 33--55, 2011.

\bibitem{I-MMSE}
D.~Guo, S.~Shamai, and S.~Verd{\'u}, ``Mutual information and minimum
  mean-square error in {G}aussian channels,'' \emph{IEEE Trans. Inf. Theory},
  vol.~51, no.~4, pp. 1261--1282, April 2005.

\bibitem{bobkov2015entropy}
S.~Bobkov and G.~P. Chistyakov, ``Entropy power inequality for the r{\'e}nyi
  entropy.'' \emph{IEEE Trans. Inf. Theory}, vol.~61, no.~2, pp. 708--714,
  2015.

\bibitem{aczel1966lectures}
J.~Acz{\'e}l, \emph{Lectures on functional equations and their
  applications}.\hskip 1em plus 0.5em minus 0.4em\relax Academic press, 1966,
  vol.~19.

\bibitem{resnick2013probability}
S.~I. Resnick, \emph{A probability path}.\hskip 1em plus 0.5em minus
  0.4em\relax Springer Science \& Business Media, 2013.

\bibitem{poor2013introduction}
H.~V. Poor, \emph{An introduction to signal detection and estimation}.\hskip
  1em plus 0.5em minus 0.4em\relax Springer Science \& Business Media, 2013.

\bibitem{carlen1991entropy}
E.~Carlen and A.~Soffer, ``Entropy production by block variable summation and
  central limit theorems,'' \emph{Communications in mathematical physics}, vol.
  140, no.~2, pp. 339--371, 1991.

\bibitem{madiman2007generalized}
M.~Madiman and A.~Barron, ``Generalized entropy power inequalities and
  monotonicity properties of information,'' \emph{IEEE Trans. Inf. Theory},
  vol.~53, no.~7, pp. 2317--2329, 2007.

\bibitem{VerduMismatchedMSE}
S.~Verd{\'u}, ``Mismatched estimation and relative entropy,'' \emph{IEEE Trans.
  Inf. Theory}, vol.~56, no.~8, pp. 3712--3720, Aug 2010.

\bibitem{courtade2016wasserstein}
T.~A. Courtade, M.~Fathi, and A.~Pananjady, ``Wasserstein stability of the
  entropy power inequality for log-concave densities,'' \emph{arXiv preprint
  arXiv:1610.07969}, 2016.

\end{thebibliography}
\bibliographystyle{IEEEtran}

\end{document}